# 3D Corporate Tourism:
# A Concept for Innovation in Nanomaterials Engineering

**Ille C Gebeshuber***

Institute of Microengineering and Nanoelectronics (IMEN),
Universiti Kebangsaan Malaysia,
43600 UKM, Bangi, Selangor, Malaysia
and
Institute of Applied Physics,
Vienna University of Technology,
Wiedner Hauptstrasse 8–10/134, 1040 Vienna, Austria
E–mail: ille.gebeshuber@ukm.my, ille.gebeshuber@mac.com
FAX: +60 3 8925 0439
*Corresponding author

**Burhanuddin Y. Majlis**

Institute of Microengineering and Nanoelectronics (IMEN),
Universiti Kebangsaan Malaysia,
43600 UKM, Bangi, Selangor, Malaysia
FAX: +60 3 8925 0439
E–mail: burhan@vlsi.eng.ukm.my

## Abstract

Nature´s materials are complex, multi–functional, hierarchical and responsive and in most instances functionality on the nanoscale is combined with performance on the macroscale. Materials engineers have just started to produce complex nanomaterials. Biomimicry and biomimetics deal with knowledge transfer from nature to technology. Inspired by the "Biomimicry and Design Workshops" by the US based Biomimicry Guild, "3D Corporate Tourism", a solution–based approach to innovation in nanomaterials research, is proposed. The three main pillars of this integrated concept are *discover*, *develop* and *design*. Biologists, research and development engineers as well as designers jointly work in an environment with high inspirational potential and construct first prototypes and designs on site. This joint approach yields new links, networks and collaborations between communities of thinkers in different countries in order to stimulate and enhance creative and application–oriented problem solving for society.

## Keywords

# Biographical Notes

Ille C. Gebeshuber is expert in Nanotechnology and Biomimetics. Since 2009 she has been Contract Professor at the Institute of Microengineering and Nanoelectronics (IMEN), Universiti Kebangsaan Malaysia (UKM). Her permanent professorship affiliation is with the Vienna University of Technology (Austria) where she is habilitated for experimental physics. She is cofounder of the Vienna–based Center of Excellence for Biomimetics. Prof. Gebeshuber is Associate Editor of the UK based Journal of Mechnical Engineering Science (IMechE), Editorial Board Member of various scientific journals and editor of a book on biomimetics by Springer Scientific Publishing. She is highly active in Science Outreach and Board Member of UKM´s Permata Pintar programm, identifying and promoting Malaysian geniuses at early age.

Burhanuddin Y. Majlis is professor of microelectronics at Universiti Kebangsaan Malaysia. He is senior member of the Institution of Electrical and Electronics Engineer (IEEE) and was Chairman of IEEE Electron Devices Malaysia Chapter from 1994 to 2006. He is the founder chaiman of Malaysia Nanotechnology Association that was established in 2007. He initiated research in microfabrication and microsensors at UKM in 1995 and has also initiated research in GaAs technology with Telekom Malaysia. In 2001 he stared research in MEMS with substantial research funding of US$10 million from the Malaysian Ministry of Science, Technology and Innovation. His current interests are design and fabrication of MEMS sensors, RFMEMS, BioMEMS and microenergy. He has published four text books in electronics and one book on Integrated Circuits Fabrication Technology for undergraduate courses and more than 300 academic research papers. Now he is the founder director of IMEN.

## Acknowledgements

The Austrian Society for the Advancement of Plant Sciences funded part of this work via the Biomimetics Pilot Project "BioScreen".

Living in the tropics and exposure to high species diversity at frequent excursions to the tropical rainforests is a highly inspirational way to do biomimetics. Profs. F. Aumayr, H. Störi and G. Badurek from the Vienna University of Technology are acknowledged for enabling ICG three years of research in the inspiring environment in Malaysia.

# 3D Corporate Tourism:
# A Concept for Innovation in Nanomaterials Engineering

**Abstract** Nature´s materials are complex, multi–functional, hierarchical and responsive and in most instances functionality on the nanoscale is combined with performance on the macroscale. Materials engineers have just started to produce complex nanomaterials. Biomimicry and biomimetics deal with knowledge transfer from nature to technology. Inspired by the "Biomimicry and Design Workshops" and the "Biomimicry Innovation Method" by the US based Biomimicry Guild, "3D Corporate Tourism", a solution–based approach to innovation in nanomaterials research, is proposed. The three main pillars of this integrated concept are *discover*, *develop* and *design.* Biologists, research and development engineers as well as designers jointly work in an environment with high inspirational potential and construct first prototypes and designs on site. This joint approach yields new links, networks and collaborations between communities of thinkers in different countries in order to stimulate and enhance creative and application–oriented problem solving for society.

**Keywords** bioinspiration, biomimetics, production methods, natural materials, biological engineering, biomimicry, innovation, nanomaterials, nanotechnology, nanofabrication, design, corporate tourism

**1. Introduction**

Natural materials are complex on various levels of hierarchy, from the nanometer length scale via the microscale to the macroscale (Aizenberg et al., 2005; Fratzl and Weinkamer, 2007). Materials engineers have just started to produce complex nanomaterials (Lao, Wen, and Ren, 2002; Cao, 2007; Tao et al., 2007), and there is still a long way to go to reach the natural "best practice" examples in terms of precision, functionality and efficiency of production (Ryadnov, 2009). Due to nanotechnology´s inherently inter– and transdisciplinary nature, increasingly, collaborations across fields prove successful (Gebeshuber, 2007; Gebeshuber and Drack, 2008; Porter and Youtie, 2009; Gebeshuber et al., 2010). Such inter– and transdisciplinary approaches are highly useful for innovation.

For companies it is not interesting if the material is a nanomaterial or not, what is interesting for them is that the material performs as intended and can be produced at reasonable costs. Therefore, a solution–based approach to innovation in materials research seems to be more rewarding and beneficial, for the companies as well as for the scientists and engineers who develop and produce such materials on the prototype level. Bioinspirations seems to be especially rewarding for innovation regarding nanomaterials, since most of the materials in nature are hierarchical, with important functionalities on the nanoscale.

One example for this is recent research on biosynthesis of nanoparticles: Various microbes such as bacteria, yeast and fungi produce inorganic nanostructures and metallic nanoparticles with properties similar to chemically–synthesised materials, while exercising strict control over size, shape and composition of the particles (see e.g., Sastry et al. 2003; Mandal et al. 2006; Gericke and Pinches 2006a; Gericke and Pinches 2006b). Examples include the formation of magnetic nanoparticles by magnetotactic bacteria (Roh et al., 2001), the production of silver nanoparticles by the bacterium *Pseudomonas stutzeri* (Klaus et al., 1999) and the fungi *Fusarium oxysporum* (Senapati et al. 2004) and *Verticillium sp.* (Mohanpuria, Rana and Yadav, 2008), synthesis of nano–scale, semi–conducting CdS crystals in the yeast *Schizosaccharomyces pombe* (Kowshik et al., 2003), the synthesis of gold nanoparticles (Gericke and Pinches, 2006b) and the formation of palladium nanoparticles using sulphate reducing bacteria in the presence of an exogenous electron donor (Yong et al., 2002). Various "best practices" for novel nanomaterials can be found in nature. Abstraction and understanding of their design principles is the prerequisite for successful transfer to technology.

Nature´s materials are like a treasure box for the nanomaterials engineer. There are functional materials, hierarchical materials, functional gradient materials, responsive materials, materials with an expiration date (i.e., they only work for a certain amount of time) and biodegradable materials (Meyers et al., 2007). One intriguing example for the exqusite materials engineering in nature is biomineralisation. More than 60 different biominerals such as are produced by organisms, including carbonates, phosphates, arsenates, chlorides and silfides (Weiner and Dove, 2003; Sigel, Sigel and Sigel, 2008; Behrens and Baeuerlein, 2009). A biomineral consists of an inorganic part (about 97 vol%), and about 3 vol% of organic matrix. The way of production of biominerals varies substantially from current materials synthesis: In the production of biominerals, properties such as shape, size and crystallinity as well as isotopic and trace element compositions are controlled by the organic matrix (Weiner and Dove, 2003). Figure 1 shows an example for biomineralised $SiO_2$ in the diatom *Ellerbeckia arenaria*. Diatoms are unicellular microalgae with a cell wall consisting of a siliceous skeleton enveloped by a thin organic case (Round, Crawford and Mann, 1990). The cell walls of each diatom form a pillbox-like shell consisting of two parts that fit within each other. These microorganisms vary greatly in shape, ranging from box-shaped to cylindrical; they can be symmetrical as well as asymmetrical and exhibit an amazing diversity of nanostructured frameworks. These biogenic hydrated silica structures have elaborate shapes, interlocking devices, and, in some cases, hinged structures (Gebeshuber and Crawford, 2006), making them interesting for emerging micro- and nanoelectromechanical systems (MEMS and NEMS, Gebeshuber et al., 2009).

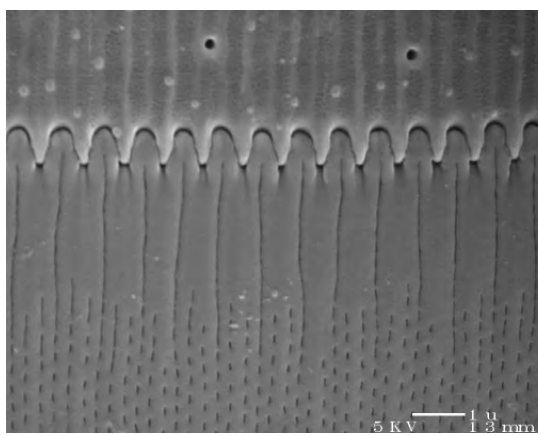

Figure 1. Biomineralised amorphous silica from the diatom *Ellerbeckia arenaria*. Scalebar: 1μm.

This article aims at mapping new frontiers in emerging and developing technology areas in nanomaterials research and innovation. A novel way to foster and promote innovative thinking in the sciences is introduced, considering the need for synergy and collaboration between biology, engineering, materials science and nanotechnology rather than segmentation and isolation: Supported by specially trained biologists, research and development engineers as well as designers apply the Biomimicry Innovation Method (© Biomimicry Guild, Helena, MT, USA 2008; Gebeshuber et al., 2009) in an environment with high inspirational potential and discover, develop and design complex nanomaterials inspired by nature. Directly at the site of this research, first prototypes and designs are constructed, and first detailed investigations take place. The three main pillars of this approach are *discover*, *develop* and *design* – the integrated concept is therefore termed "3D Corporate Tourism" (for logo, see Figure 2, for basic concept see Figure 3). This concept has been inspired by the "Biomimicry and Design Workshops" (offered for one week per year, location: rainforest in Peru or in Costa Rica) by the US based Biomimicry Guild. Companies such as Boeing, Colgate–Palmolive, General Electric, Levi's, NASA, Nike and Procter and Gamble have already used their services. Janine Benyus, the founder of the Biomicry Guild, states: "*When we take designers and engineers and architects on these workshops – these are people who everyday are inventing. They usually get their inspiration by looking at other human inventions. We got them outside, where they were surrounded in Costa Rica by the most amazingly sustainable system that one can imagine, and so much variety. Each of those organisms was solving amazing challenges, but they were solving them in very, very different ways. ..... There was somebody there who was working on packaging, and for them it was a revelation because here's this packaging, which was the exoskeleton of the beetle. It breathes, it signals, it creates its beautiful color without toxic chemicals, it's waterproof, it's manufactured in a completely nontoxic way, and it's abrasion–resistant. All the things they would love to have in packaging, and yet it's made of one material that is completely recyclable.*" (Benyus, 2009).

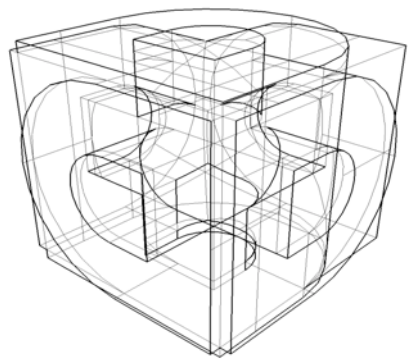

Figure 2. Logo of "3D Corporate Tourism".

With "3D Corporate Tourism", the successful concept of the "Biomimicry and Design Workshops" is developed further into a complete niche tourism concept.

The outcome of such a joint effort are – besides the research results, developments and designs – new links, networks and collaborations between communities of thinkers in different countries in order to stimulate and enhance creative and application–oriented problem solving for society.

## 2. Materials and Methods

*2.1. Method*

Research and development engineers as well as designers from companies work at "3D Corporate Tourism" headquarters on the three pillars of the concept: discover, develop, design. The Biomimicry Innovation Method (BIM, © Biomimicry Guild, Helena, MT, USA 2008; Gebeshuber et al., 2009) is applied to identify high–potential biological systems, processes and materials that shall inspire novel outcomes. BIM is an innovation method that seeks sustainable solutions by emulating Nature's time–tested patterns and strategies. The goal is to create products, processes, and policies – new ways of living – that are well adapted to life on earth over the long haul.

The steps in BIM are as follows: Identify function, biologise the question, find nature's best practices and generate product ideas.

**Identify function:** The biologists distil challenges posed by engineers/natural scientists/architects and/or designers to their functional essence. In the case of nanomaterial development and innovation, self–repairing structures, tough and yet flexible materials, photoactive materials such as luminescent fungi, and the natural recycling system that is at work in the rainforest (the waste of one species is nurturing another species, a concept that has as "Waste to Wealth" already entered current science and economy (Kathiravale and Muhd Yunus, 2008)).

**Biologise the question:** In the next step, these functions are translated into biological questions such as "How does Nature provide strong and yet flexible materials?" or "How does Nature convert energy?" or "How does Nature self–repair structures?" The basic question is "What would Nature do here?"

**Find Nature's best practices:** Screens of the relevant literature in scientific databases as well as entering a highly inspiring environment with the biologised questions in mind (task–oriented visit) are used to obtain a compendium of how plants, animals and ecosystems solve the challenges in question. The inspiring environments should preferably be habitats with high species diversity, e.g., rain forests or coral reefs. Thereby a compendium of how plants, animals and ecosystems solve the specific challenge is obtained.

**Generate process/product ideas:** From these best practices (according to the Biomimicry Guild, 90% of them are usually new to clients) ideas for cost effective, innovative, life–friendly and sustainable products and processes are generated. First designs and investigations are already turned into prototypes right at the site.

In general the projects are structured in three phases (discover, develop and design, Figures 3 and 4).

*2.1.1. Discovery Phase*

The discovery phase aims to define the key elements that are needed for a solution and to approach nature for inspiration, here the support of experts will be important as the way to look at nature and to isolate segments with solution potential has to be acquired. Here nature will be observed with a 'function view' (What works like we need it?).

*2.1.2. Development Phase*

The development phase is probably the most important phase and comprises all activities that are undertaken to convert the solution of nature into human technology. Here top–engineering and prototyping support will be crucial. Nature will be screened if needed with 'process view' (How is the solution achieved in detail? Which process steps are there?).

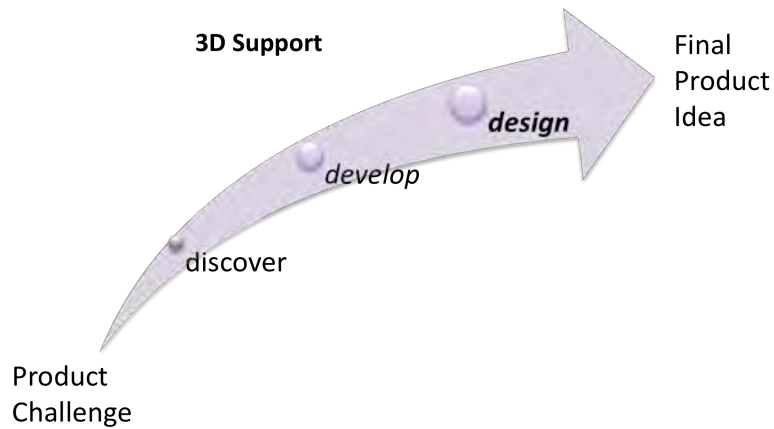

Figure 3. The three pillars of the "3D Corporate Tourism" concept, starting from the product challenge and resulting in the final product ideas.

To achieve the best possible results it is suggested to insert a normal holiday (one or two weeks with family in a Rainforest Resort) between the development and the design phase; that way the team members can relax and rethink the project results before finalizing the design. This period is also a kind of time reserve in case the prototyping cannot be finished in time, which is very likely the case for more complicated designs.

*2.1.3. Design Phase*

The design phase focuses on the optimization and final design of the defined solution. The final design will be further improved by a second inspiration loop with nature ('design view': Why do the components of nature work together this well? Where is the detail optimization?). Here experts and designers will help the team to prepare a top–quality presentation for their corporate headquarters.

*2.2. Project Team*

The project team consists of the visiting engineers, developers and designers from the respective companies, and specifically trained biologists and engineers trained in biomimetics from the host country. The visitors either arrive with a specific problem they would like to deal with, or the theme(s) of the project are determined on site, after initial introduction to the respective approaches, joint discussions and brainstorming sessions, and inspirational task–oriented visits to the rainforest. There is also the possibility of MSc and PhD students to join the projects and contribute their knowledge and ideas. Information flow is free between the members of the individual projects, and intellectual property rights are secured by standard documents that are signed at the beginning of the project.

*2.3. Environment*

The "3D Corporate Tourism" headquarters are small lodges located directly inside the rainforest or close to coral reefs. They have sufficient standard for the clientele from the companies and are equipped with amenities such as CAD machines, state of the art communication facilities, scientific and patent databases and prototyping machinery.

This distinguishes them from the currently available rainforest biology field stations. The lodges are utilised by one project team at one time, and joint social activities shall foster discussion in the free time.

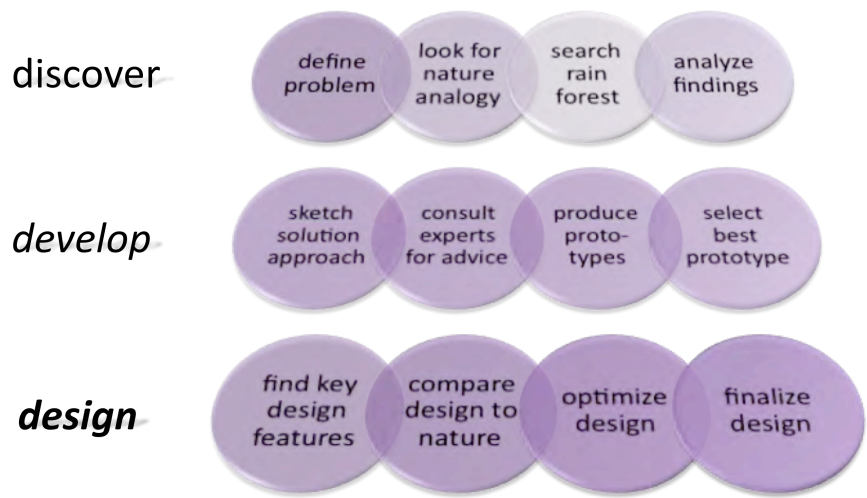

Figure 4. Details of the three main pillars of the "3D Corporate Tourism" concept.

## 3. Results and Discussion

The rainforest serves as nanotechnology lab, and combined with the "3D Corporate Tourism" headquarters provides a complete solution for work from the concept phase to first prototypes (Figure 5). In this way, collaborations between local scientists and engineers with the international clientele are initiated, bidirectional knowledge transfer takes place and new knowledge is generated.

The high species variety in the rainforest, with nature´s "best practices" everywhere aids to relate structure with function in natural materials, structures and processes and helps to increase awareness about the natural resources surrounding us. With the concept of "3D Corporate Tourism" the potential of the virgin rainforests is used in a sustainable way, without exploiting the natural resources or removing anything else from the jungle apart from ideas.

In this way, the value of the virgin forests is increasing in the minds of policy makers and threshold countries have the opportunity to contribute highly valued inputs to the international research and development elite, as well as train their local experts in very important future technologies. The possibility to perform first investigations directly on–site, and subsequent deeper and more detailed investigations at the home institution fosters collaborations and results in synergistic effects across borders.

## 4. Conclusion

The "3D Tourism" concept seems to be especially interesting for development of novel functional nanomaterials and contributes to overcoming the first two of the three gaps between the world of ideas, inventors, innovators and investors as well as the market as introduced by Gebeshuber, Gruber and Drack, 2009 for accelerated scientific and technological breakthroughs to improve the human condition (Figure 5).

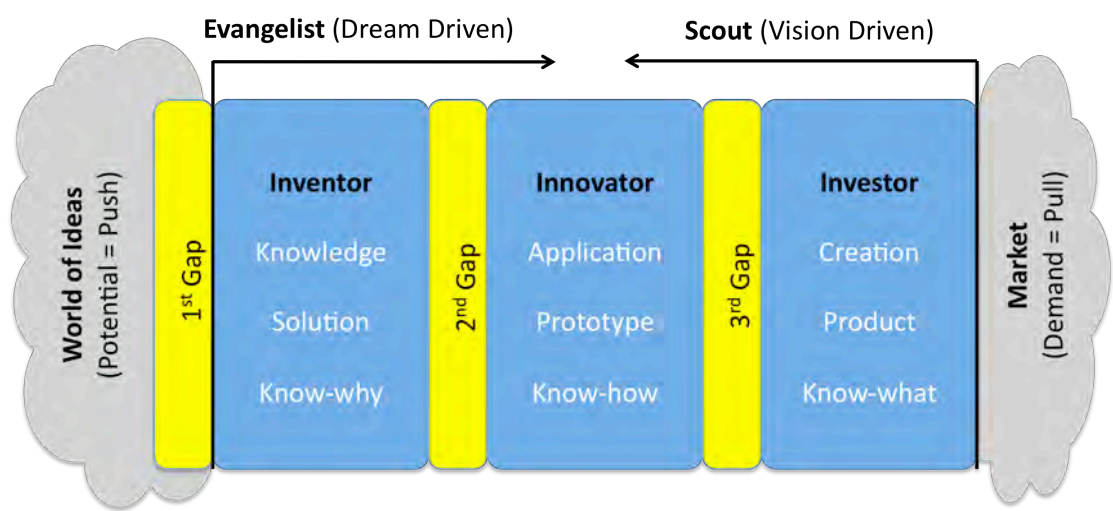

Figure 5. The Three–Gaps–Theory as proposed by Gebeshuber, Gruber and Drack, 2009. Image © 2009 Professional Engineering Publishing, UK. Image reproduced with permission.